\documentclass[12pt,reprint,aps,groupedaddress,nofootinbib,prd,twocolumn]{revtex4-1}

\usepackage[english]{babel}
\usepackage[utf8]{inputenc}
\usepackage{amsmath}
\usepackage{mathbbol}
\usepackage{amssymb}
\usepackage{bbold}
\usepackage{graphicx,amsfonts}
\usepackage{epsfig}
\usepackage{xcolor}
\definecolor{crimsonglory}{rgb}{0.75,0.0,0.2}
\usepackage[colorlinks=true,
linkcolor=crimsonglory,
urlcolor=crimsonglory,
citecolor=crimsonglory]{hyperref}
\usepackage{capt-of}
\usepackage{bm}
\usepackage{mathrsfs}
\usepackage{enumerate}
\usepackage{amsthm}
\usepackage{bbm}
\usepackage{comment}
\usepackage{appendix}
\usepackage{physics}
\usepackage{url}
\usepackage{upgreek}

\newcommand{\be}{\begin{equation}}
	\newcommand{\ee}{\end{equation}}
\newcommand{\beq}{\begin{equation}}
	\newcommand{\eeq}{\end{equation}}
\newcommand{\bea}{\begin{eqnarray}}
	\newcommand{\eea}{\end{eqnarray}}
\newcommand{\bit}{\begin{itemize}}
	\newcommand{\eit}{\end{itemize}}
\newcommand{\ben}{\begin{enumerate}}
	\newcommand{\een}{\end{enumerate}}

\begin{document}
	
\preprint{APS/123-QED}

\title{Hidden heavy flavor tetraquarks in the Born-Oppenheimer approximation}

\author{Bowen Kang}
\affiliation{College of Physics, Chengdu University of Technology, Chengdu 610059, China}

\author{Xi Xia}
\affiliation{College of Physics, Chengdu University of Technology, Chengdu 610059, China}

\author{Tao Guo}
\email[Corresponding author:]{guot17@tsinghua.org.cn}
\affiliation{College of Physics, Chengdu University of Technology, Chengdu 610059, China}


\date{\today}%
	
	\begin{abstract}
		
		The Born-Oppenheimer approximation is one of the very successful tools for solving the hydrogen atom problem. The experimental discovery of hidden heavy flavor tetraquarks, $Q\bar{Q}q \bar{q}$ ($Q=c,b$ and $q=u,d,s$), provides great possibilities for the hydrogen-bond-like structure of the Quantum Chromodynamics version. 
		In this work, considering that the colors of $Q\bar{Q}$ and $q\bar{q}$ are both $8$,  the tetraquark $Q\bar{Q}q \bar{q}$ system is formed by color coupling $8\otimes8 \rightarrow 1$.
		In order to study the mass splitting caused by the color-spin hyperfine interaction, the color-spin basis vectors of the $S$-wave tetraquark states are appropriately constructed.
		Then we use the Born-Oppenheimer approximation to calculate the mass spectra of the $S$-wave hidden heavy flavor tetraquark states. 
		The results show that some of the hidden heavy flavor exotic hadrons discovered experimentally can be well explained as this type of hydrogen-bond-like tetraquark structure. 
		In addition, some candidates for tetraquark bound states are predicted and may be compact tetraquark states.
		
	\end{abstract}
	
	\maketitle
	
\section{Introduction}

Quantum chromodynamics (QCD) provides a rich framework for understanding hadrons, extending beyond the conventional mesons (one quark, one antiquark) and baryons (three quarks) to encompass exotic configurations such as multiquark states, hybrids, and glueballs \cite{Godfrey:1985xj,Klempt:2007cp,Lebed:2016hpi}. 
Among the multiquark states, the study of tetraquark states (composed of two quarks and two antiquarks) have received significant attention \cite{Jaffe:1976ig,Jaffe:1976ih,Lodha:2024bwn}.
Notably, certain hidden heavy flavor tetraquarks, $Q\bar{Q}q \bar{q}$ ($Q=c,b$ and $q=u,d,s$), can be interpreted as analogues of the hydrogen bonds within QCD version.
This allows the interaction between heavy quark-antiquark pairs and light quarks can be treated using the Born-Oppenheimer (BO) approximation \cite{Esposito:2013fma,Braaten:2014qka,Maiani:2019cwl,Maiani:2019lpu,Mutuk:2024vzv,Berwein:2024ztx,Berwein:2024ztx,Germani:2025qhg}. 
Based on the fact that heavy quarks are much heavier than light quarks, we can decouple the heavy quark dynamics from the light quark dynamics. 
This strategy makes it possible to systematically study the mass spectra and structures of the hidden heavy flavor tetraquark states.

In 2003, the Belle Collaboration reported the famous narrow charmonium-like state $X(3872)$ in the $\pi^+\pi^- J/\psi$ mass spectrum of the exclusive decay process $B^\pm \rightarrow K^{\pm}\pi^+\pi^-J/\psi$ \cite{Belle:2003nnu}.
The discovery of the $X(3872)$ confirmed the existence of exotic hadron states and opened up a new direction for hadron physics research.
On the theoretical side, there are various explanations for the nature of the $X(3872)$, such as the tetraquark state \cite{Maiani:2004vq,Matheus:2006xi,Chen:2016qju,Wang:2023sii}, molecular state \cite{Wong:2003xk,AlFiky:2005jd,Liu:2008fh,Guo:2013sya}, $P$-wave charmonium \cite{Barnes:2003vb,Bali:2011dc,Mohler:2012na,Aliev:2014doa}, and threshold effects \cite{Bugg:2004rk,Hanhart:2007yq,Guo:2014iya}.
Subsequently, more hidden heavy flavor exotic hadrons were discovered, many of which were considered to be candidates for tetraquark states and attracted much attention from particle physicists, such as $X(4140)$ \cite{CDF:2009jgo,Padmanath:2015era,Luo:2017eub}, $Y(3940)$ \cite{Belle:2004lle,Ebert:2005nc,Lebed:2016yvr}, $Z_b(10610)$ and $Z_b(10650)$ \cite{Belle:2011aa,Guo:2011gu,Leskovec:2019ioa}; see reviews for details \cite{Karliner:2017qhf,Guo:2017jvc,Brambilla:2019esw,Chen:2022asf}. 
So far, although the understanding of these exotic hadrons has not yet reached a consensus and is still under intense scientific debates, they deserve further investigation as candidates for  tetraquark states.

Recently, there is increasing experimental evidence proving the existence of possible hidden heavy flavor tetraquark states.
In 2020, the first candidate for a charged hidden-charm tetraquark state with strangeness, $Z_{cs}(3985)^-$, was discovered in the processes of $e^+e^- \rightarrow K^+D_s^- D^{\ast 0}$ and $K^+D_s^{\ast -} D^{0}$ by the BESIII collaboration \cite{BESIII:2020qkh}.
Its non-zero charge characteristic enables it to be clearly distinguished from hybrid or charmonium.
Theoretically, $Z_{cs}(3985)^-$ is generally interpreted as tetraquark state $cs\bar{c}\bar{u}$ with a quantum number assignment $J^P=1^+$ \cite{Meng:2020ihj,Karliner:2021qok,Giron:2021sla,Guo:2022crh,Liu:2024uxn}.
Later in 2021 the LHCb collaboration reported two hidden-charm exotic structures in the decay $B^+ \rightarrow J/\psi\phi K^+$, referred to as $Z_{cs}(4000)^+$ and $Z_{cs}(4220)^+$ \cite{LHCb:2021uow}. 
Their quark composition is thought to be $cu\bar{c}\bar{s}$, and their spin-parity quantum quantum numbers are determined to be $J^P=1^+$ and $1^+$ (or $1^-$), respectively \cite{LHCb:2021uow}.
However, theoretical studies suggest that the $Z_{cs}(4000)^+$ and $Z_{cs}(4220)^+$ may be candidates for charmonia \cite{A:2023bxv,Bokade:2024tge}, tetraquarks \cite{Li:2023wug,Vogt:2024fky,Yu:2024ljg}, or molecules \cite{Yang:2020nrt,Chen:2023iee}.
It is worth noting that the BESIII collaboration recently observed a new $Z_{cs}(4123)^-$ in the processes of $e^+e^- \rightarrow K^+ D_s^{\ast -} D^{\ast 0} + c.c.^\ast$, and its quantum number assignment may be $J^P=1^+$ \cite{BESIII:2022vxd}. 
The discovery of the $Z_{cs}(4123)^-$ may add a new member to the family of charged hidden-charm tetraquark with strangeness.
In 2023, A near-threshold peaking structure $X(3960)$ is observed in the $D_s^+D_s^-$ invariant mass spectrum by  the LHCb collaboration, and the quantum number assignment $J^{PC}=0^{++}$ is favored \cite{LHCb:2022aki}.
Research shows that the nature of $X(3960)$ may also be a tetraquark $cs\bar{c}\bar{s}$ \cite{Badalian:2023qyi,Hoffer:2024alv,Chen:2025rxl}.
Obviously, these exotic hadrons all have the common components $Q\bar{Q}q \bar{q}$, and figuring out their nature will help us better understand QCD.

In this work, we apply the BO approximation to calculate the mass spectra of tetraquark $Q\bar{Q}q \bar{q}$. 
The heavy $Q\bar{Q}$ forms a color octet, which combines with a light $q \bar{q}$ octet to produce a color-singlet tetraquark, that is, the color coupling is represented as $8\otimes8 \rightarrow 1$.
Given the same amount of momentum, the heavy quarks move much more slowly than the light quarks.
The light quark (or antiquark) interactions can be treated as perturbations to the dominant heavy-quark potential, allowing the Schrödinger equation for the heavy quark separation to be solved numerically. 
The calculation of diquark-antidiquark paradigm was extended to tetraquark system. 
Using $X(3872)$ as input, other possible tetraquark states were studied.

This paper is organized as follows. 
In Sec. \ref{S2}, We give a brief introduction of the tetraquark Hamiltonian based on the BO approximation. 
We obtain the BO potential using the perturbation method and construct the basis vector wave functions of the tetraquark system $Q\bar{Q}q \bar{q}$.
In Sec. \ref{S3}, we determine the model parameters, such as the effective masses of quarks and the color coupling strengths. 
The mass spectra and decay modes of various hidden heavy tetraquark states are calculated and discussed in this section. 
We summarize in Sec. \ref{S4}.
The nature units $c = \hbar = 1$ are used throughout.

\section{The Born-Oppenheimer theoretical model \label{S2}}
\subsection{Hamiltonian in the Born-Oppenheimer approximation}
For the hidden heavy flavor tetraquark system, it is theoretically difficult to achieve a direct analytical solution. We can use BO approximation in non-relativistic quantum mechanics to handle the degrees of freedom of heavy and light quarks separately. The Hamiltonian of tetraquark system reads
\begin{eqnarray}\nonumber
	{\cal H} &=& \sum_{i=1}^{4} \left( \frac{p_i^2}{2m_i} \right) + V(\bm{x}_{A,B}) \\
	         &+& V_I(\bm{x}_{A,B}, \bm{x}_{1,2})+ V_{\text{conf}},\label{eq1}
\end{eqnarray}
where $\bm{x}_{A,B}$ and $\bm{x}_{1,2}$ represent the coordinates of heavy quarks and light quarks respectively, $m_i$ represents the effective mass of the $i$-th constituent quark.
$\sum_{i=1}^{4} \left(\frac{p_i^2}{2m_i} \right)$ is defined as the kinetic energy of the quarks. 
$V(\bm{x}_{A,B})$ describes the interaction between the heavy quarks. 
$V_I(\bm{x}_{A,B},\bm{x}_{1,2})$ describes the general interactions involving light-heavy and light-light quarks. 
$V_{\text{conf}}$ is defined as the color confinement part of the tetraquark states.

If we focus only on the $S$-wave tetraquark states, the spatial degrees of freedom can be encoded into the model parameters.
Considering the energy level splitting caused by the color-spin interaction, the Hamiltonian for the $S$-wave tetraquark states be simplified as \cite{Maiani:2004vq,Maiani:2019lpu}
\begin{equation}
H = \sum_{i=1}^{4} m_i + E + H_{ss},\label{eq2}
\end{equation}
with
\begin{equation}
    H_{ss} = \sum_{i<j} 2\kappa_{ij} (\bm{s}_i \cdot \bm{s}_j),\label{eq3}
\end{equation}
where $E$ is the energy eigenvalue of the tetraquark system obtained based on the BO approximation.
The coefficient $\kappa_{ij}$ is inversely proportional to the product of the masses of the $i$-th and $j$-th constituent quarks, and directly proportional to the color coupling strength of the corresponding quark pair.
The Hamiltonian $H_{ss}$ is the color-spin hyperfine interaction.

\subsection{Born-Oppenheimer potential}
In the BO approximation, with the positions of the heavy quarks assumed to be fixed, we first consider the Schrödinger equation for the light quarks\cite{Maiani:2019lpu}
\begin{equation}
\begin{split}
\left[ \left( \sum_{\text{light}} \frac{p_i^2}{2m} \right) 
+ V_I(\bm{x}_{A,B},\bm{x}_{1,2}) \right]f = \mathcal{E}(\bm{x}_{A,B}) f,\label{eq4}
\end{split}
\end{equation}
where $f = f(\bm{x}_{A,B},\bm{x}_{1,2})$ is the eigenfunction for the light quarks. $\mathcal{E}(\bm{x}_{A,B})$ is the eigenvalue dependent on the positions of the heavy quarks. 
The total wavefunction of tetraquark system can be written as
\begin{equation}
\Psi = \psi(\bm{x}_{A,B}) f(\bm{x}_{A,B},\bm{x}_{1,2}), \label{eq5}
\end{equation}
where $\psi$ describes the wavefunction of heavy quarks.
We let the Hamiltonian Eq.(\ref{eq1}) act on the wave function Eq.(\ref{eq5}).
Since the wavefunction $f$ of light quarks is very sensitive to changes in the position of heavy quarks, $\nabla_i f$ cannot be ignored.
But the wavefunction $\psi$ of heavy quarks changes slowly with respect to the positions of the heavy quarks, $\nabla_i \psi$ can be approximated as zero.
For the kinetic term, it can be approximated as
\begin{equation}
\nabla_i^2 \Psi \approx \left( f \nabla_i^2 \psi + \psi \nabla_i^2 f \right).\label{eq7}
\end{equation}
Now we can rewrite the Schrödinger equation of heavy quarks as
\begin{equation}
\left( \sum_{\text{heavy}} \frac{p_i^2}{2M} +  V_{\text{BO}}(\bm{x}_{A,B})\right) \psi 
= E \psi,\label{eq8}
\end{equation}
with
\begin{equation}
V_{\text{BO}}(\bm{x}_{A,B}) = V(\bm{x}_{A,B}) + \mathcal{E}(\bm{x}_{A,B}) + V_{\text{conf}}.\label{eq9}
\end{equation}

Next, we analyze the specific form of $V_{\text{BO}}$ potential.
In a hadron system, the color coupling between any quark and quark (or antiquark) can be expressed as
\begin{equation}
V(r) = \lambda_{q_1 q_2}(\bm{R}) \frac{\alpha_s}{r},\label{eq10}
\end{equation}
with the Casimir operator
\begin{equation}
	\lambda_{q_1 q_2}(\bm{R}) = \frac{1}{2} \left[ C(\bm{R}) - C(q_1) - C(q_2) \right],\label{eq13}
\end{equation}
where $r$ is the relative distance between the two quarks, and $\alpha_s$ is the QCD running coupling constant. 
The notation $\bm{R}$ is the color representation of any pair of particles, $q_1$ and $q_2$ represent the irreducible representations of the quark.
In the SU(3) group representation, we can get $C(3) = 4/3$, $C(6) = 10/3$, $C(1) = 0$, $C(8) = 3$, $C(\bar{q}) = C(q)$, and $C(\bm{\bar{R}}) = C(\bm{R})$.

For hidden heavy flavor tetraquarks, the color couplings of the two heavy quarks can be either $1$ or $8$.
In the first case, however, the interaction between $ Q\bar{Q} $ and $ q\bar{q} $ pairs occurs through the exchange of color singlets. 
This is similar to the nucleon-nucleon interaction, which manifests itself as the configuration of hadronic molecular states.
Therefore,we exclude the scenario where $ Q\bar{Q} $ couples in the color state $1$. 
In our calculation, we only consider the coupling of quark pairs both ${Q\bar{Q}}$ and ${q\bar{q}}$ with color state $8$.

In order to derive $\mathcal{E}$, we need to solve Schrödinger Eq.\eqref{eq4}.
Given that the impact of heavy quarks on light quarks is non-negligible, the hidden heavy flavour tetraquark state can be considered as a hydrogen-like molecular system of QCD version.
Then, we can extend the calculation methods for hydrogen molecules to study the tetraquark $ Q\bar{Q}q\bar{q} $.
Specifically, a light quark is paired exclusively with one heavy quark (or antiquark) forming two symmetrical orbits. 
To analyze the interaction between the orbit $\bar{Q}q$ or $Q \bar{q}$, we introduce the Cornell potential of non-relativistic quark model,
\begin{equation}
    V_{\bar{Q}q}(r) = -\frac{4}{3} \frac{\alpha_s}{r} + \sigma r + V_0,\label{eq16}
\end{equation}
where $\sigma$ is the string tensor strength of QCD and is proportional to $|\lambda|$. The parameter $V_0 \simeq -0.315\,\text{GeV}$ can be obtained by fitting the quarkonium.
For the orbit $Qq$ or $\bar{Q} \bar{q}$, we modify the interaction as follows
\begin{equation}
    V_{Qq}(r) = -\frac{1}{3} \frac{\alpha_s}{r} + \frac{1}{4} \sigma r + V_0.\label{eq18}
\end{equation}
The radial variation wave function is given by
\begin{equation}
R(r) = \frac{A^{3/2}}{\sqrt{4\pi}} e^{-Ar},\label{eq19}
\end{equation}
where $A$ is the variational parameter.
Substituting the trial wave function and interaction into the Hamiltonian, we can obtain the ground state energy ($\langle H \rangle_{\text{min}}$) and the corresponding wave function ($\zeta$ or $\xi$) of a heavy-light quark pair by using variational calculation.

We consider as nonperturbative process for the orbits of two non-interacting heavy-light quark pairs.
Due to the symmetry of the tetraquark system, the wave function of light quarks can be expressed as
\begin{equation}
f = \zeta(\bm{\rho})\xi(\bm{\eta}) = R(|\bm{x}_1 - \bm{x}_A|)R(|\bm{x}_2 - \bm{x}_B|),\label{eq20}
\end{equation}
and the ground state energy corresponding to the wave function reads
\begin{equation}
E_{Qq \leftrightarrow \bar{Q} \bar{q}} = 2 \langle H \rangle_{\text{min}}.\label{eq21}
\end{equation}
For the orbit formed by the light quark-antiquark pair, we regard it as the perturbation term. Then, the perturbation Hamiltonian is
\begin{eqnarray}\nonumber
H_{p} = &-& \frac{7}{6}\alpha_s \left( \frac{1}{|\bm{x}_1 - \bm{x}_B|} + \frac{1}{|\bm{x}_2 - \bm{x}_A|} \right) \\
&+& \frac{1}{6}\alpha_s \frac{1}{|\bm{x}_1 - \bm{x}_2|}.\label{eq22}
\end{eqnarray}

For simplicity, the perturbative energy is written by
\begin{equation}
\Delta E = \langle f | H_{p} | f \rangle,\label{eq23}
\end{equation}
By calculation, $\Delta E$ is written as
\begin{equation}
\Delta E = -\frac{7}{6}\alpha_s 2 I_1(r_{AB}) + \frac{1}{6}\alpha_s I_2(r_{AB}),\label{eq24}
\end{equation}
where $r_{AB}=|\bm{x}_B-\bm{x}_A|$. Functions $ I_1 $ and $ I_2 $ are
\begin{equation}
    I_1(r_{AB}) = \int \text{d}^3 \rho |\psi(\rho)|^2 \frac{1}{|\bm{\rho} - \bm{r}_{AB}|},\label{eq25}
\end{equation}
\begin{equation}
I_2(r_{AB}) = \int \text{d}^3\rho\, \text{d}^3\eta |\psi(\rho)|^2 |\phi(\eta)|^2 \frac{1}{|\bm{\rho} - \bm{\eta}|}.\label{eq26}
\end{equation}
Now, we can obtain the final energy as
\begin{equation}
    \mathcal{E}(\bm{x}_{A,B}) = E_{Qq \leftrightarrow \bar{Q} \bar{q}} + \Delta E.\label{eq27}
\end{equation}

For the hidden heavy flavor tetraquark system, the overall color coupling needs to ensure color singlet.
We need to include a linear string tension to simulate the confinement effect\cite{Bruschini:2020voj,Andreev:2022cax}.
Considering the tetraquark configuration $|(Q\bar{Q})_{\boldsymbol{8}}(q\bar{q})_{\boldsymbol{8}}\rangle_{\boldsymbol{1}}$, the linearly increasing confinement term reads \cite{Maiani:2019lpu}
\begin{equation}
    V_{\text{conf}} = \sigma \times (r - R_0) \times \theta(r - R_0).\label{eq28}
\end{equation}
$R_0$ is the relative distance between quarks within a hadron.
In general, $R_0$ can be treated as an adjustable free parameter and can be obtained by fitting the hadron spectrum.
Therefore, the Born-Oppenheimer potential can be expressed as
\begin{eqnarray}\nonumber
V_{\text{BO}}(r_{AB}) &=& \frac{1}{6}\alpha_S \frac{1}{r_{AB}} + E_{Qq \leftrightarrow \bar{Q} \bar{q}} + \Delta E\\
                 &+& \sigma \times (r_{AB} - R_0) \times \theta(r_{AB} - R_0).\label{eq30}
\end{eqnarray}
Substituting Eq.(\ref{eq30}) into Eq.(\ref{eq8}), we can obtain the binding energy $E$ of the hidden heavy flavor tetraquark state under the BO approximation.

\subsection{Wave function}

In order to study the energy level splitting caused by color-spin hyperfine interaction, we can construct the basis vector wave functions of the tetraquark state.
In color space, the tetraquark state is divided into meson-meson configuration $|(Q\bar{Q})(q\bar{q})\rangle$ (or $|(Q\bar{q})(q\bar{Q})\rangle$) and diquark-antidiquark configuration $|(Qq)(\bar{Q}\bar{q})\rangle$. And these two configurations can be connected by Fierz transformation.
In this calculation, we focus only on the diquark-antidiquark configuration.
Obviously, the total spin of all possible $S$-wave tetraquark states be 0, 1, and 2.
For the scalar tetraquark states with quantum number $J^{P} = 0^{+}$, there are two basis vectors
\begin{equation}
	\begin{aligned}
		&|0^{+(+)} \rangle_1 = |0_{Qq}, 0_{\bar{Q}\bar{q}}; J = 0\rangle, \\
		&|0^{+(+)} \rangle_2 = |1_{Qq}, 1_{\bar{Q}\bar{q}}; J = 0\rangle.\label{eq31}
	\end{aligned}
\end{equation}
The symbol $(+/-)$ represents charge conjugation when a tetraquark state is in flavor symmetry. 
And the positive and negative signs in quantum numbers have the same meaning below, unless otherwise stated.

For the axial vector tetraquark states with quantum number $J^P = 1^+$, there are three basis vectors
\begin{equation}
\begin{aligned}
    &|A\rangle = |0_{Qq}, 1_{\bar{Q}\bar{q}}; J = 1\rangle, \\
    &|B\rangle = |1_{Qq}, 0_{\bar{Q}\bar{q}}; J = 1\rangle, \\
    &|C\rangle = |1_{Qq}, 1_{\bar{Q}\bar{q}}; J = 1\rangle.\label{eq32}
\end{aligned}
\end{equation}
In charge conjugation, considering the symmetric and antisymmetric combinations of states $|A\rangle$ and $|B\rangle$, we obtain a $C$-odd and a $C$-even state.
And the charge conjugation of $|C\rangle$ is odd. 
Therefore, for the hidden heavy flavor tetraquark states with $J^P=1^+$, the basis vectors with definite charge conjugation can be rewritten as
\begin{equation}
\begin{aligned}
    &|1^{+(+)}\rangle = \frac{1}{\sqrt{2}} (|A\rangle + |B\rangle), \\
    &|1^{+(-)}\rangle_1 = \frac{1}{\sqrt{2}} (|A\rangle - |B\rangle), \\
    &|1^{+(-)}\rangle_2 = |C\rangle.\label{eq33}
\end{aligned}
\end{equation}

For the $J^{P} = 2^{+}$ state, the basis vector is
\begin{equation}
|2^{+(+)}\rangle = |1_{Qq}, 1_{\bar{Q}\bar{q}}; J = 2\rangle.\label{eq34}
\end{equation}

Now we have constructed the basis vector wave functions of the hidden heavy flavor tetraquark system based on the diquark-antidiquark configuration.
Substituting these basis vectors into Eq.(\ref{eq2}), we can obtain the corresponding Hamiltonian matrices.
Then, by diagonalizing the Hamiltonian matrices, we can obtain the mass spectra and corresponding wave functions of the tetraquark states.
Finally, considering the characteristics of the meson-meson configuration, we analyze possible decay channels of these calculated tetraquark states.

\section{ Results and discussion \label{S3}}

\subsection{Model parameters}

In the specific numerical calculation, we first need to determine the model parameters.
By fitting the experimental observations and lattice calculations of heavy flavor conventional hadrons, we extract the effective masses of the constituent quarks: $m_n = 308$ MeV, $m_s = 484$ MeV, $m_c = 1667$ MeV, and $m_b = 5005$ MeV.
Here $n$ represents $u$ and $d$ quarks.
Without loss of generality, we take $\alpha_s$ and $\sigma$ based on the charmonium, bottomonium, and bottom-charm mesons: $\alpha_s(2m_c) = 0.30$, $\alpha_s(m_c + m_b) = 0.24$, $\alpha_s(2m_b) = 0.21$, and $\sigma = 0.15$ GeV$^2$.
In addition, other model parameters required for our calculation are listed in the Tables \ref{tab:spin_spin_couplings}-\ref{tab:spin_spin_baryons}.

\begin{table}[h]
    \centering
    \caption{Parameter $\kappa$ (in MeV) for quark-antiquark systems. And these parameters are derived from fitting the $S$-wave mesons.}
    \begin{ruledtabular}
    \begin{tabular}{lcccccccccc}
        & $n\bar{n}$ & $s\bar{n}$ & $s\bar{s}$ & $c\bar{n}$ & $c\bar{s}$ & $c\bar{c}$ & $b\bar{n}$ & $b\bar{s}$ & $b\bar{c}$& $b\bar{b}$ \\
        \hline
        $(\kappa_{ij})_{\boldsymbol{0}}$ &  315 & 195 & 121 & 70 & 72 & 59 &  23 & 24 & 20 & 30 \\
    \end{tabular}
    \label{tab:spin_spin_couplings}
   \end{ruledtabular}
\end{table}
\begin{table}[h]
    \centering
    \caption{Parameter $\kappa$ (in MeV) for quark-quark (or antiquark-antiquark) systems. And these parameters are derived from fitting the $S$-wave baryons.}
    \begin{ruledtabular}
    \begin{tabular}{lcccccc}
        & $nn$ & $sn$ & $cn$ & $cs$ & $bn$ & $bs$ \\
       \hline
       $(\kappa_{ij})_{\bar{\boldsymbol{3}}}$ & 103 & 64 & 22 & 25 & 6.6 & 7.5\\
    \end{tabular}
    \label{tab:spin_spin_baryons}
    \end{ruledtabular}
\end{table}

So far, since the $s\bar{s}$ meson with quantum number $J^P=0^-$ and the $b\bar{c}$ meson with quantum number $J^P=1^-$ have not been observed experimentally, the parameters $\kappa_{s\bar{s}}$ and $\kappa_{b\bar{c}}$ can be extracted by using a similar method as in Ref. \cite{Maiani:2004vq}.
It is worth emphasizing that these parameters have a certain symmetry, namely $\kappa_{ij}=\kappa_{ji}$.

\begin{figure}[h]
	\centering
	\includegraphics[width=0.46\textwidth,height=0.28\textwidth]{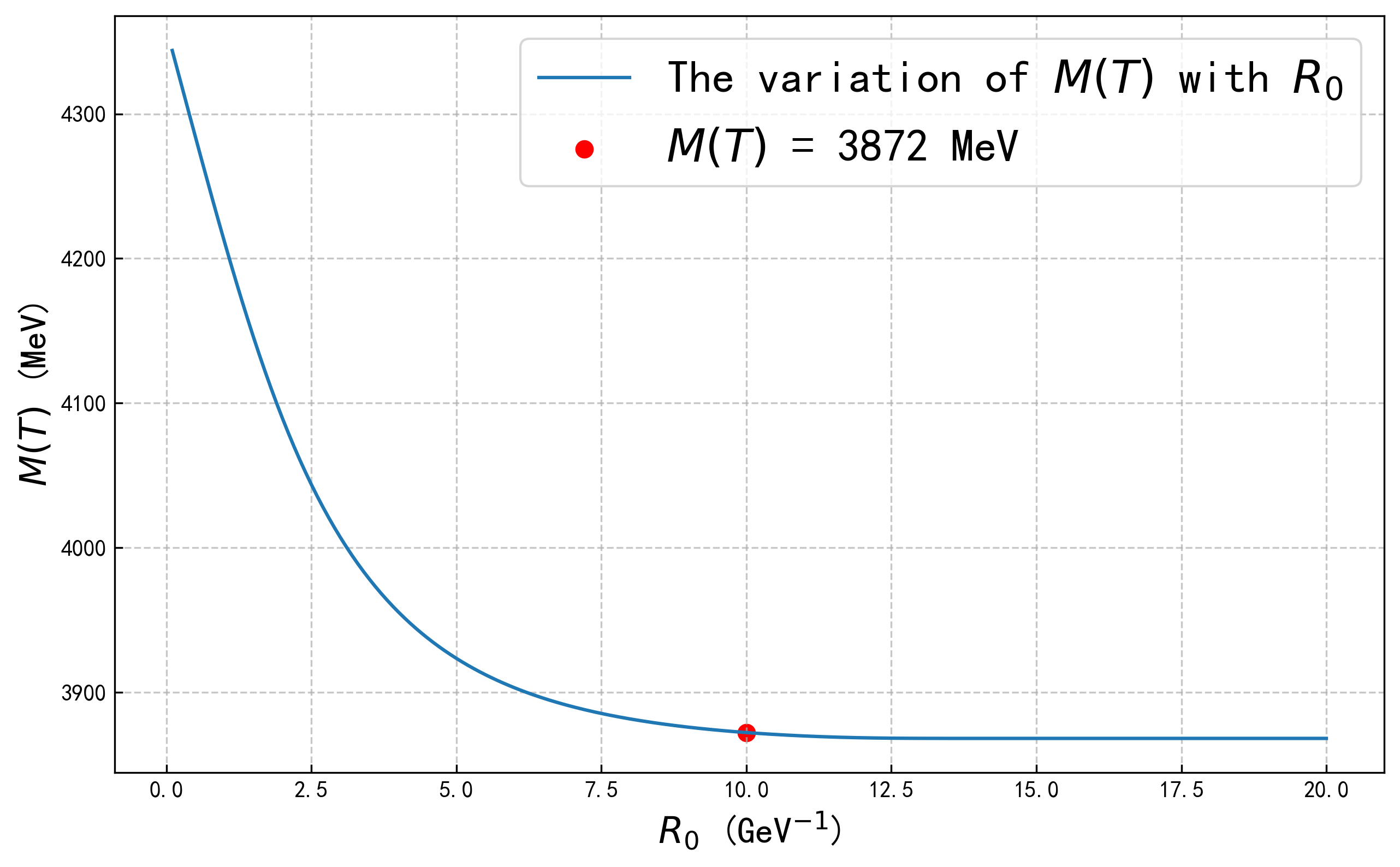}
	\vspace{-0.5cm}
	\caption{For the tetraquark state $c\bar{c}n\bar{n}$ with quantum number $J^{PC}=1^{++}$, the mass spectrum of the ground state decreases with increasing parameter $R_0$.}
	\label{fig:mass_decay}
\end{figure}

To solve the mass spectra of the hidden heavy flavor tetraquark states, we choose $X(3872)$ as input to determine $R_0$.
Assuming that $X(3872)$ is a tetraquark state $c\bar{c}n\bar{n}$ with quantum number $J^{PC}=1^{++}$, we can substitute the basis vector $|1^{++}\rangle$ into the Hamiltonian Eq. (\ref{eq2}) to obtain the corresponding mass spectrum
\begin{equation}
    M(T) = 2(m_c + m_q) + E + H_{ss},\label{eq41}
\end{equation}
where $T$ represents the tetraquark state $c\bar{c}n\bar{n}$.
Now we plot the mass $M(T)$ as $R_0$ changes, see Figure \ref{fig:mass_decay}.
It can be found that the mass $M(T)$ decreases as $R_0$ increases.
It is noteworthy that when $R_0 \ge$ 10 $\text{GeV}^{-1}$, the mass spectrum of the tetraquark state $c\bar{c}n\bar{n}$ tends to be constant. 
This indicates that the confinement between $c\bar{c}$ disappears and the tetraquark state decays into two open-charm mesons.
Besides, we calculated the mass spectra of the tetraquark state $c\bar{c}n\bar{n}$ at different $R_0$, see Table \ref{tab:mass_values_three_line}.
Among them, when $R_0 = 10 \, \text{GeV}^{-1}$, the mass spectrum of the tetraquark state $ c\bar{c}n\bar{n} $ with quantum number $1^{++}$ is 3872 MeV.

\begin{table}[h]
    \centering
    \caption{Mass spectra of tetraquark $c\bar{c}n\bar{n}$ at varying $R_0$.}
    \begin{ruledtabular}
    \begin{tabular}{lccc c}
        $R_0$ (GeV$^{-1}$) & \multicolumn{4}{c}{Mass (MeV)} \\
        & $M(0^{+(+)})$ & $M(1^{+(-)})$ & $M(1^{+(+)})$ & $M(2^{+(+)})$ \\
        \hline
        $R_0 = 6$ & 3792, 3898 & 3785, 3913 & 3903 & 3982  \\
        $R_0 = 8$ & 3770, 3876 & 3763, 3891 & 3881 & 3960 \\
        $R_0 = 10$ & 3761, 3867 & 3754, 3882 & \fbox{3872} & 3951 \\
        $R_0 = 12$ & 3757, 3863 & 3750, 3878 & 3868 & 3947 \\
    \end{tabular}
    \end{ruledtabular}
    \label{tab:mass_values_three_line}
\end{table}

\begin{figure*}[htbp]
  \centering
  \begin{minipage}{0.32\textwidth}
    \centering
    \includegraphics[width=\linewidth]{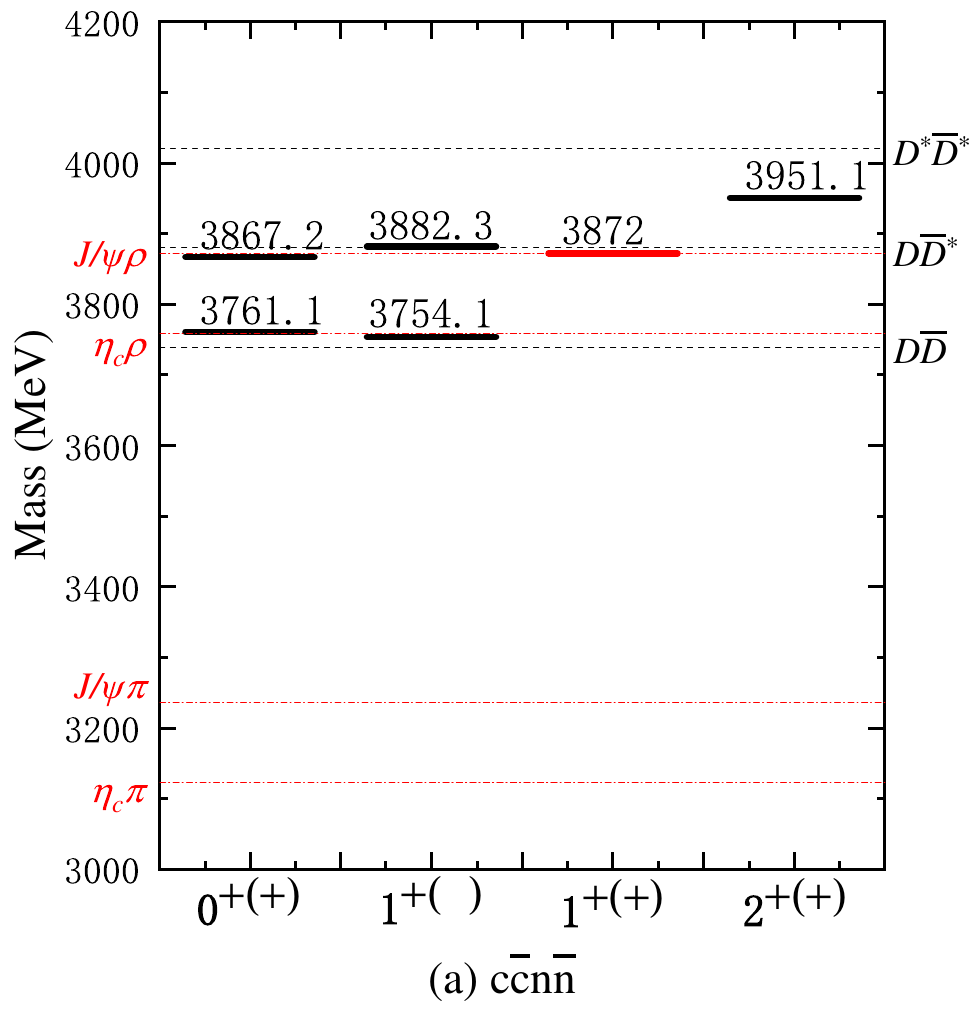}
   
  \end{minipage}\hfill
  \begin{minipage}{0.32\textwidth}
    \centering
    \includegraphics[width=\linewidth]{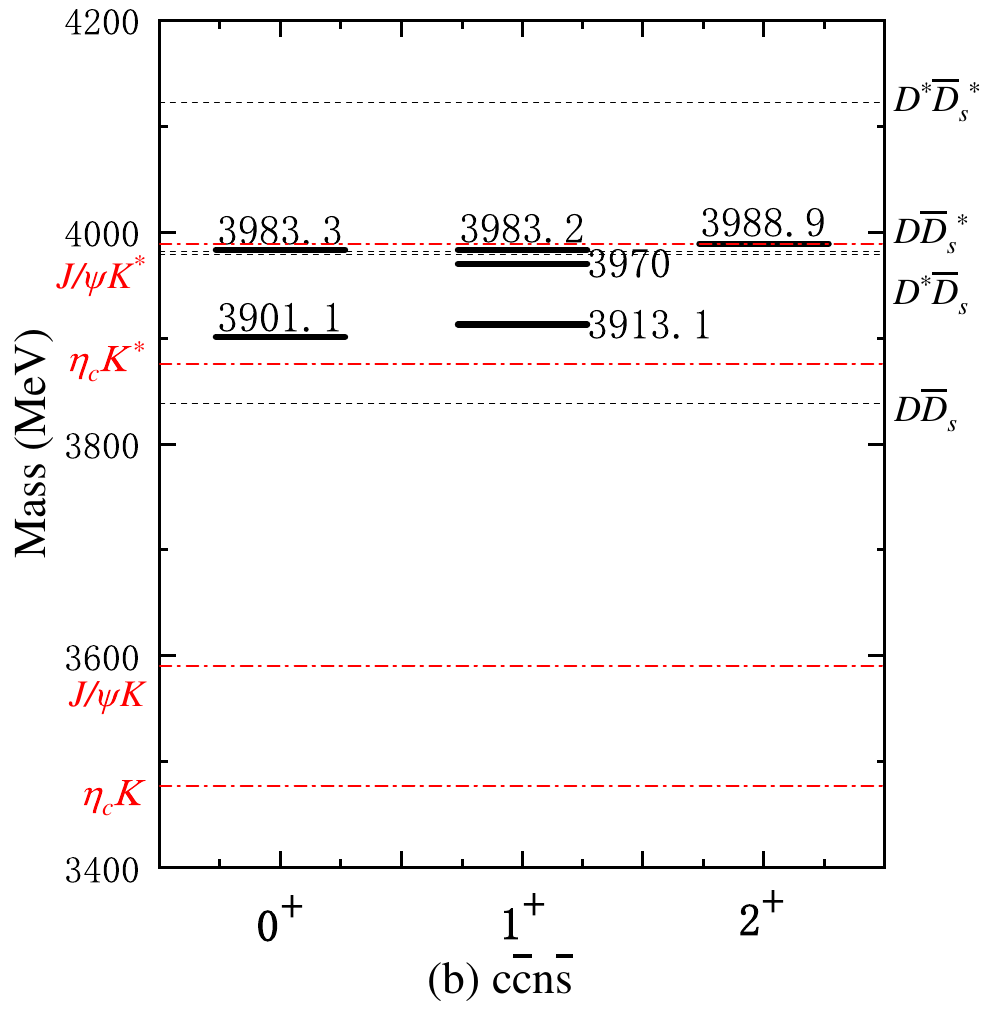}
    
  \end{minipage}\hfill
  \begin{minipage}{0.32\textwidth}
    \centering
    \includegraphics[width=\linewidth]{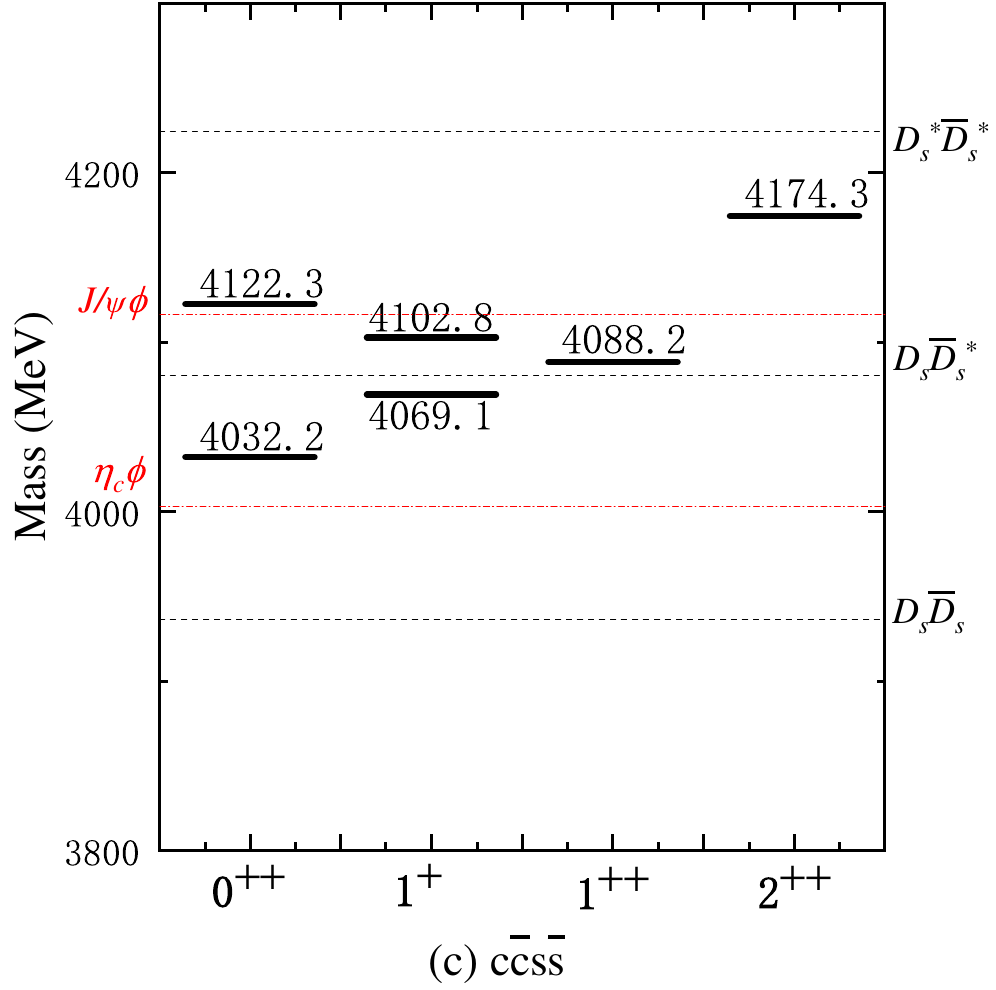}
    
  \end{minipage}
  \caption{The mass spectra of the $ S $-wave hidden-charm tetraquark states: (a) $ c\bar{c}n\bar{n} $, (b) $ c\bar{c}n\bar{s} $, and (c) $ c\bar{c}s\bar{s} $. The black dashed lines and red dash-dotted lines represent the corresponding meson-meson thresholds.
}
  \label{fig:three_images_1}
\end{figure*}

\begin{figure*}[htbp]
  \centering
  \begin{minipage}{0.32\textwidth}
    \centering
    \includegraphics[width=\linewidth]{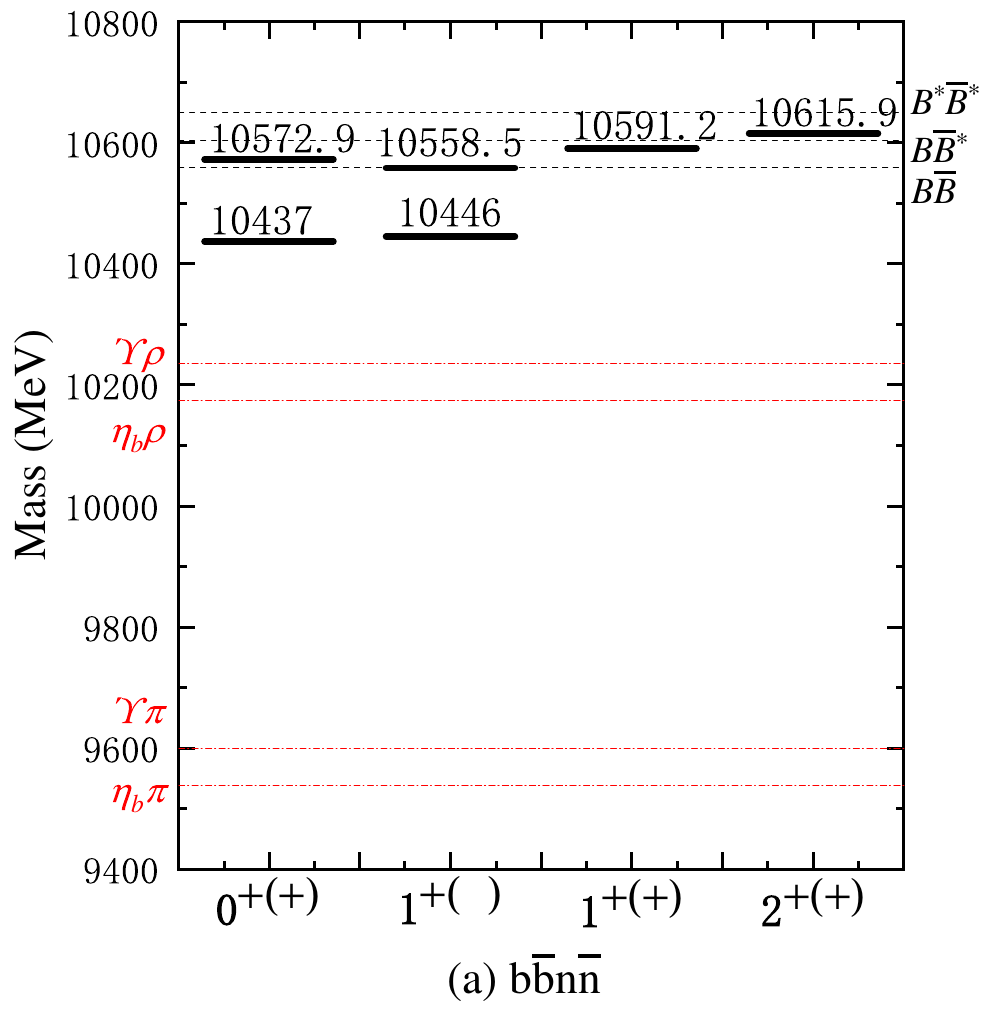}
   
  \end{minipage}\hfill
  \begin{minipage}{0.32\textwidth}
    \centering
    \includegraphics[width=\linewidth]{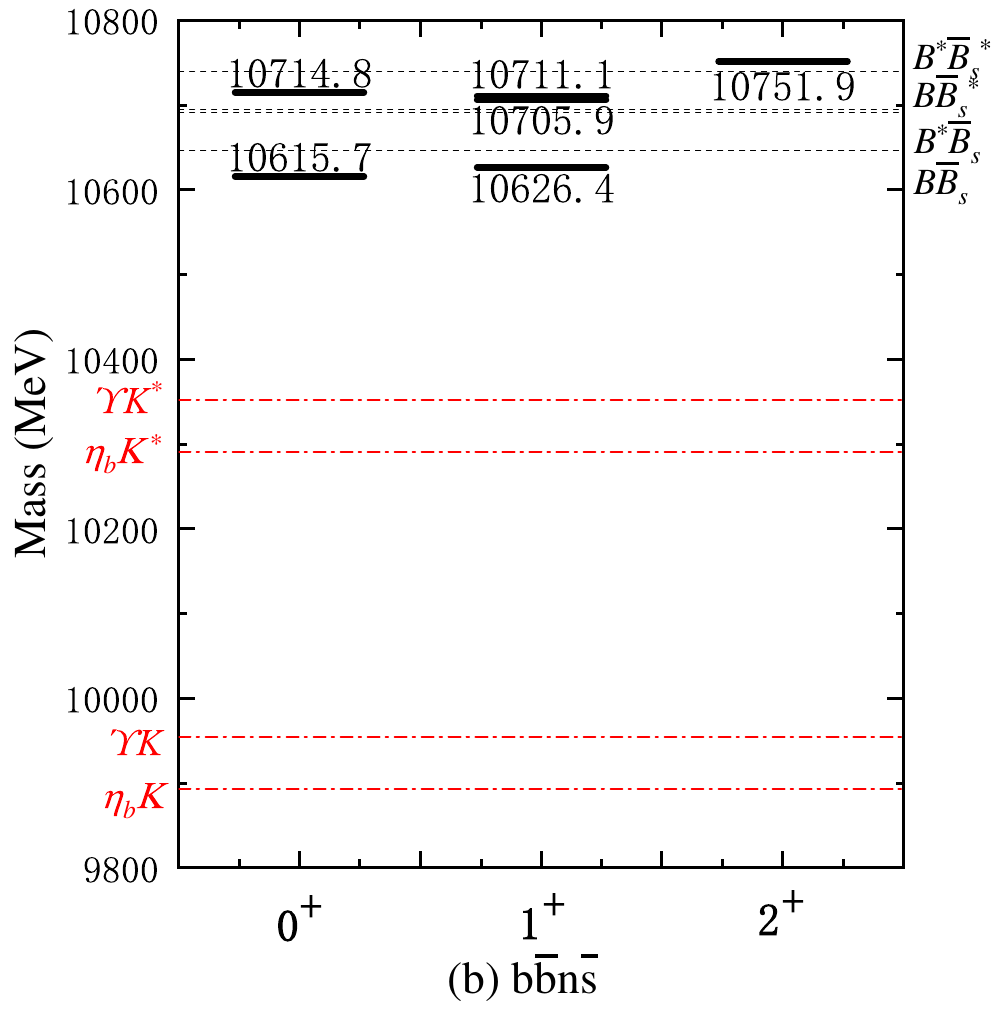}
   
  \end{minipage}\hfill
  \begin{minipage}{0.32\textwidth}
    \centering
    \includegraphics[width=\linewidth]{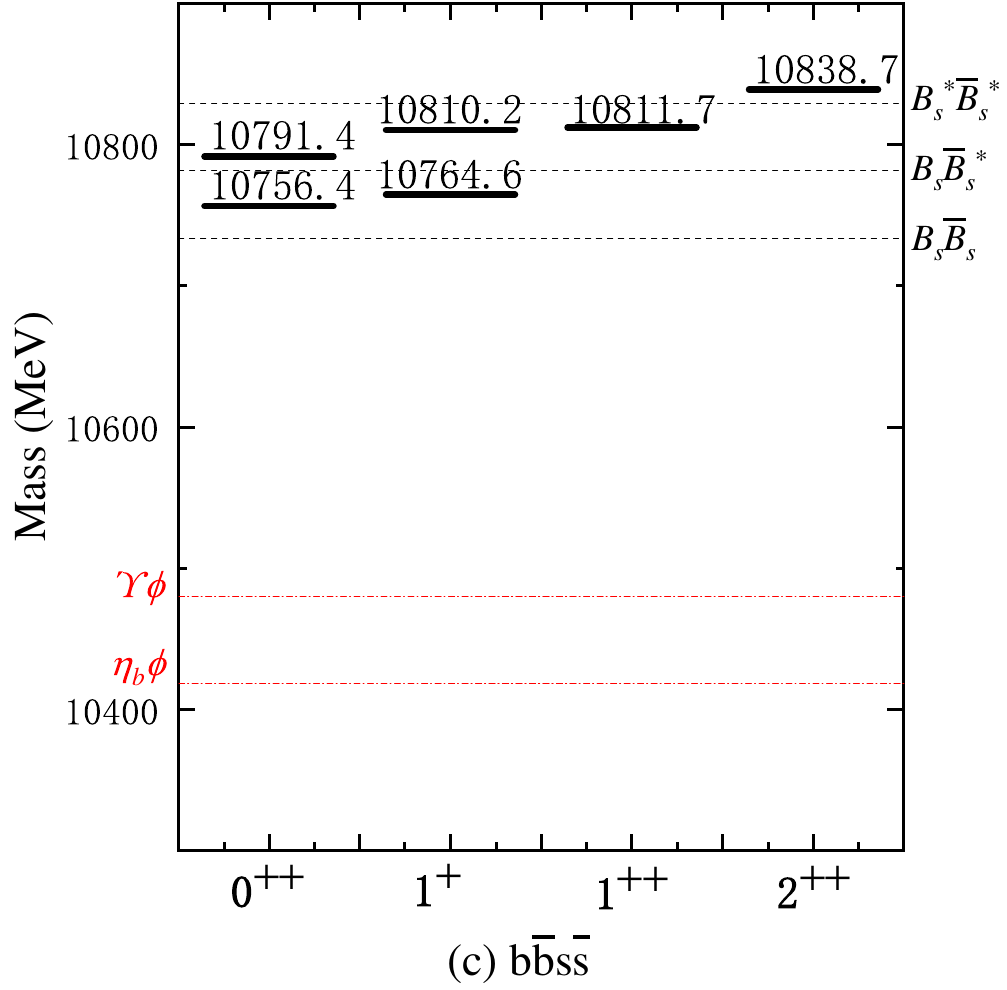}
   
  \end{minipage}
  \caption{The mass spectra of the $ S $-wave hidden-bottom tetraquark states: (a) $ b\bar{b}n\bar{n} $, (b) $ b\bar{b}n\bar{s} $, and (c) $ b\bar{b}s\bar{s} $. The black dashed lines and red dash-dotted lines represent all possible meson-meson thresholds.}
  \label{fig:three_images_2}
\end{figure*}

\subsection{Hidden-charm tetraquark states with $ c\bar{c}n\bar{n} $, $ c\bar{c}n\bar{s} $, and $ c\bar{c}s\bar{s} $ configurations}

In this section, we systematically investigate the mass spectra of the $S$-wave hidden-charm tetraquark states by properly considering the BO approximation and the color-spin hyperfine interaction.
And the quantum number assignments of these tetraquark states are $J^{P(C)}=0^{+(+)}$, $1^{+(\pm)}$, and $2^{+(+)}$.
The model parameters and specific calculation scheme have been explained in the previous section.
We take $R_0=10$ GeV$^{-1}$, and the mass spectra of the tetraquark states $ c\bar{c}n\bar{n} $, $ c\bar{c}n\bar{s} $, and $ c\bar{c}s\bar{s} $ are plotted in Figure \ref{fig:three_images_1}, respectively.
For ease of comparison, the corresponding meson-meson thresholds are also plotted in these subfigures.
Besides, we have also calculated the mass spectra of the tetraquarks $c\bar{c}n\bar{s}$ and $c\bar{c}s\bar{s}$ at varying $R_0$ values, as shown in Tables \ref{tab:mass_values_cc_qbar_s}-\ref{tab:mass_values_cc_sbar_s}.

\subsubsection{The $ c\bar{c}n\bar{n} $ configuration}

For the tetraquark state $c\bar{c}n\bar{n}$ with different quantum numbers, the mass spectra and the corresponding meson-meson thresholds are exhibited in Figure \ref{fig:three_images_1}(a).
In this tetraquark system, if the two light quarks are of the same flavor, the isospin of $c\bar{c}n\bar{n}$ is $I=0$ and it is electrically neutral.
If one of the two light quarks is a $u$ quark and the other is a $d$ quark, the isospin of $c\bar{c}n\bar{n}$ is $I=1$ and it is a charged particle.
At the same time, the possible final state decay channels of these calculated tetraquark states are analyzed according to the quantum numbers and meson thresholds.

There are two $S$-wave hidden-charm tetraquark states with $J^{P(C)}=0^{+(+)}$, whose masses are 3761.1 and 3867.2 MeV respectively.
The state with the lowest mass, 3761.1 MeV, lies above the $ D\bar{D} $ and $\eta_c\pi$ thresholds.
Considering the conservation of quantum numbers ($J$, $P$, and $C$), this state can decay naturally into mesons $D\bar{D}$ or $\eta_c\pi$.
It shows that this state may be observed in the invariant mass spectrum of mesons $D\bar{D}$ or $\eta_c\pi$.
In addition, in this article, it should be emphasized that we only analyze the $S$-wave decay process of the tetraquark state and ignore the higher-order $P$-wave and $D$-wave decay modes.
The reason is that the higher-order decay channels are generally small and are not the main decay modes.
The state, 3867.2 MeV, is also above the $ \eta_c\pi $ and $ D\bar{D} $ thresholds but slightly below the $ J/\psi \rho $ threshold. 
Therefore, this state is allowed to decay into mesons $ \eta_c\pi $, $ D\bar{D} $, or $ J/\psi \rho $.
It can be considered as a candidate for $X(3915)$.

There are three $ S $-wave hidden-charm states with $ J^P = 1^+ $.
Their masses are 3754.1, 3872.0, and 3882.3 MeV respectively.
this type of tetraquark states can be classified into two categories based on their charge conjugation $ C $-parity. 
For quantum number $ J^{PC} = 1^{+-} $, the lowest state, 3754.1 MeV, is slightly below the $ \eta_c\rho $ threshold. 
It can be considered as a compact tetraquark state.
Normally, this state can decay into mesons $ \eta_c\pi^+\pi^- $ via an off-shell process for meson $\rho$.
In addition, this state can also decay into mesons $J/\psi \pi$.
Similarly, the state with a mass of 3882.3 MeV is close to the thresholds $ D\bar{D}^* $ and $J/\psi \rho$, suggesting that it can easily decay into mesons $ D\bar{D}^* $ or $J/\psi \rho$. 
Of course, it can also decay into mesons $ \eta_c\rho $ or $ J/\psi \pi $. 
For the state with quantum number $ J^{PC} = 1^{++} $, its mass is very close to the thresholds $ J/\psi \rho $ and $ D\bar{D}^* $.
Therefore, this state tends to decay into mesons $ J/\psi \rho $ and $ D\bar{D}^* $. And $ J/\psi \rho $ can decay naturally into mesons $ J/\psi \pi^+ \pi^- $, which is consistent with previous studies \cite{Braaten:2005ai}.
Obviously, it can be concluded that the $|X(3872)\rangle=(c\bar{c}u\bar{u}-c\bar{c}d\bar{d})/\sqrt{2}$ can be considered to be an electrically neutral tetraquark state.

There is an $ S $-wave hidden-charm tetraquark state with $ J^{P(C)} = 2^{+(+)}$. 
Its mass is 3951.1 MeV, which is above the threshold $ J/\psi \rho $.
Thus it is possible for the state to decay into mesons $ J/\psi \rho $. 
Again, this state is slightly below the threshold $ D^* \bar{D}^* $, so it may also be found in the $ D^* \bar{D}^* $ invariant mass spectrum.

\subsubsection{The $ c\bar{c}n\bar{s} $ configuration}

The mass spectra of tetraquark $ c\bar{c}n\bar{s} $ system are shown in Figure \ref{fig:three_images_1}(b). 
For comparison and decay channel analysis, the corresponding meson-meson thresholds are also presented in this figure.
For the tetraquark state $ c\bar{c}n\bar{s} $, the $ C $-parity is undetermined, so the quantum numbers of the $ S $-wave states are $ J^P = 0^+ $, $ 1^+ $, and $ 2^+ $.
In addition, we take $R_0 = 6, 8, 10$, and $12$ GeV$^{-1}$, the calculated mass spectra of the $S$-wave tetraquark state $ c\bar{c}q\bar{s} $ with varying quantum numbers are listed in Table \ref{tab:mass_values_cc_qbar_s}.

There are two $S$-wave tetraquark states with $ J^P = 0^+ $. Their masses are 3901.1 MeV and 3983.3 MeV, respectively. 
The lighter state, at 3901.1 MeV, is above the thresholds $ \eta_c K $ and $ D\bar{D}_s $, and can easily decay into mesons $ \eta_c K $ and $ D\bar{D}_s $. 
However, due to the large mass gap, the decay widths may be large, making it difficult to observe experimentally in the decay $ \eta_c K $ and $ D\bar{D}_s $ channels.
The state at 3983.3 MeV can not only decay naturally into the mesons $ \eta_c K $ and $ D\bar{D}_s $, but also decay into the mesons $ J/\psi K^* $.
Obviously, this state is close to the threshold $ J/\psi K^* $, so it is easier to be found in the $ J/\psi K^* $ invariant mass spectrum.

For $ J^P = 1^+ $,  we  find  three $ S $-wave tetraquark states with  masses of 3913.1, 3970, and 3983.2 MeV, respectively. 
The lightest state, 3913.1 MeV, is above the thresholds $ J/\psi K $ and $ \eta_c K^* $, and its decay channels are likely to include $ J/\psi K $ or $ \eta_c K^* $. 
Similarly, the state at 3970 MeV can also decay naturally into the mesons $ J/\psi K $ or $ \eta_c K^* $. 
Of course, this state is very close to the $ J/\psi K^* $ and $ D\bar{D}_s^* $ thresholds, and it is very likely to decay into the corresponding mesons.
The state at 3983.2 MeV is significantly above the thresholds $ J/\psi K $and $ \eta_c K^* $, allowing it to decay naturally into the mesons  $ J/\psi K $and $ \eta_c K^* $.
At the same time, this state is located near the thresholds $J/\psi K^* $, $ D^*\bar{D}_s $, and $ D\bar{D}_s^* $. 
There is a high probability that this state will be found in the invariant mass spectrum of $J/\psi K^* $, $ D^*\bar{D}_s $, or $ D\bar{D}_s^* $.
Obviously, these three hidden-charm tetraquark partner states with strangeness can be candidates for exotic hadrons $Z_{cs}(3985)^{-}$ , $Z_{cs}(4000)^{+}$, or $Z_{cs}(4123)^-$.

\begin{table}[h]
	\centering
	\caption{Mass spectra of tetraquark $c\bar{c}n\bar{s}$ at varying $R_0$.}
	\begin{ruledtabular}
		\begin{tabular}{lccc}
			$R_0$ (GeV$^{-1}$) & \multicolumn{3}{c}{Mass (MeV)} \\
			& $M(0^{+})$ & $M(1^{+})$ & $M(2^{+})$ \\
			\hline
			$R_0 = 6$ & 3929, 4011 & 3942, 3998, 4011 & 4017 \\
			$R_0 = 8$ & 3909, 3991 & 3922, 3978, 3991 & 3997 \\
			$R_0 = 10$ & 3901, 3983 & 3913, 3970, 3983 & 3989 \\
			$R_0 = 12$ & 3898, 3980 & 3910, 3967, 3979 & 3986 \\
		\end{tabular}
		\label{tab:mass_values_cc_qbar_s}
	\end{ruledtabular}
\end{table}

For $ J^P = 2^+ $, there is an $ S $-wave tetraquark state with a mass of 3988.9 MeV. 
Its mass spectrum lies above the $ J/\psi K^* $ threshold, allowing it to decay naturally into the mesons $ J/\psi K^* $.
Also, this state is below the threshold $ D^* \bar{D}_s^* $, allowing possible decay to the corresponding mesons via off-shell processes.
Similarly, this state can also be considered as a candidate for a compact tetraquark state.

\subsubsection{The $ c\bar{c}s\bar{s} $ configuration}

The $ S $-wave hidden-charm tetraquark $ c\bar{c}s\bar{s} $ system has a well-defined $ C $-parity. 
The possible quantum numbers are $ J^{PC} = 0^{++}$, $1^{+-}$, and $2^{++}$. 
Given $R_0 = 10$ GeV$^{-1}$, the calculated mass spectra of the tetraquark states with the $ c\bar{s}c\bar{s} $ configuration are shown in Figure \ref{fig:three_images_1}(c).
At the same time, we also show the corresponding meson-meson thresholds.
For varying $R_0$, we calculated the mass spectra of the tetraquark $ c\bar{c}s\bar{s} $ system under different quantum number assignments, see Table \ref{tab:mass_values_cc_sbar_s}.

\begin{table}[h]
	\centering
	\caption{Mass spectra of tetraquark $c\bar{c}s\bar{s}$ at varying $R_0$.}
	\begin{ruledtabular}
		\begin{tabular}{lcccc}
			$R_0$ (GeV$^{-1}$) & \multicolumn{4}{c}{Mass (MeV)} \\
			& $M(0^{++})$ & $M(1^{+-})$ & $M(1^{++})$ & $M(2^{++})$ \\
			\hline
			$R_0 = 6$ & 4060, 4150 & 4096, 4130 & 4115 & 4201 \\
			$R_0 = 8$ & 4041, 4130 & 4077, 4111 & 4096 & 4182 \\
			$R_0 = 10$ & 4032, 4122 & 4069, 4103 & 4088 & 4174 \\
			$R_0 = 12$ & 4030, 4120 & 4066, 4101 & 4086 & 4172 \\
		\end{tabular}
	\end{ruledtabular}
	\label{tab:mass_values_cc_sbar_s}
\end{table}

For $ S $-wave tetraquark state $ c\bar{c}s\bar{s} $ with $ J^{PC} = 0^{++} $, we find two states with masses of 4032.2 MeV and 4122.3 MeV, respectively. 
The lowest state, 4032.2 MeV, is above the threshold $ D_s \bar{D}_s $, so its possible decay channel is mesons $ D_s \bar{D}_s $. 
These results are in good agreement with the experimental observations of the exotic hadron $X(3960)$ by the LHCb collaboration. 
Therefore, the tetraquark state can be regarded as a good candidate for $X(3960)$.
The state at 4122.3 MeV, which lies just below the threshold $ D_s^* \bar{D}_s^* $, can naturally decay into mesons $ D_s \bar{D}_s $ and $ J/\psi \phi $.

There are two $ S $-wave tetraquark states with $ J^{PC} = 1^{+-} $, having masses of 4069.1 and 4102.8 MeV, respectively. 
The lighter state, 4069.1 MeV, is well above the $ \eta_c \phi $ threshold, allowing it to naturally decay into mesons $ \eta_c \phi $. 
The state at 4102.8 MeV, in addition to the decay mode into mesons $ \eta_c \phi $, also has a possible decay mode into mesons $ D_s \bar{D}_s^*$.
It is worth noting that this state is slightly below the threshold $J/\psi \phi$, and a corresponding off-shell decay process may be allowed.
Furthermore, there is an $ S $-wave tetraquark state with $ J^{PC} = 1^{++} $ and a mass of 4088.2 MeV. 
For this state, both $ \eta_c \phi $ and $ D_s \bar{D}_s^* $ decay modes are allowed. 
This result suggests that the tetraquark state may be a candidate for the exotic hadrons $ Y(3940) $ or $ X(4140) $.

For $ J^{PC} = 2^{++} $, we observe an $ S $-wave tetraquark state with a mass of 4174.3 MeV. 
Considering $ J^{PC} $ conservation, this tetraquark state can decay into the mesons $ J/\psi \phi $.
Of course, the channel for the tetraquark state to decay into mesons $ D_s^* \bar{D}_s^* $ cannot still be prohibited.

\subsection{Hidden-bottom tetraquark states with $ b\bar{b}n\bar{n} $, $ b\bar{b}n\bar{s} $, and $ b\bar{b}s\bar{s} $ configurations}

\begin{table}[h]
	\centering
	\caption{Mass spectra of tetraquark $b\bar{b}n\bar{n}$ at varying $R_0$.}
	\begin{ruledtabular}
		{\tabcolsep 0.00325in
			\begin{tabular}{lcccc}
				$R_0$ (GeV$^{-1}$) & \multicolumn{4}{c}{Mass (MeV)} \\
				& $M(0^{+(+)})$ & $M(1^{+(-)})$ & $M(1^{+(+)})$ & $M(2^{+(+)})$ \\
				\hline
				$R_0 = 6$  & 10583, 10447 & 10456, 10598 & 10601 & 10625 \\
				$R_0 = 8$  & 10575, 10439 & 10448, 10591 & 10593 & 10618 \\
				$R_0 = 10$ & 10573, 10437 & 10446, 10559 & 10591 & 10616 \\
				$R_0 = 12$ & 10572, 10436 & 10445, 10588 & 10591 & 10615 \\
			\end{tabular}
		}
		\label{tab:mass_values_bb_qbar_q}
	\end{ruledtabular}
\end{table}
\begin{table}[h]
	\centering
	\caption{Mass spectra of tetraquark $b\bar{b}n\bar{s}$ at varying $R_0$.}
	\begin{ruledtabular}
		\begin{tabular}{lccc}
			$R_0$ (GeV$^{-1}$) & \multicolumn{3}{c}{Mass (MeV)} \\
			& $M(0^{+})$ & $M(1^{+})$ & $M(2^{+})$  \\
			\hline
			$R_0 = 6$  & 10637, 10736 & 10648, 10727, 10732 & 10773 \\
			$R_0 = 8$  & 10622, 10721 & 10633, 10712, 10717 & 10758 \\
			$R_0 = 10$ & 10616, 10715 & 10626, 10706, 10711 & 10752 \\
			$R_0 = 12$ & 10613, 10712 & 10624, 10704, 10709 & 10750 \\
		\end{tabular}
		\label{tab:mass_values_bb_qbar_s}
	\end{ruledtabular}
\end{table}
\begin{table}[h]
	\centering
	\caption{Mass spectra of tetraquark $b\bar{b}s\bar{s}$ at varying $R_0$.}
	\begin{ruledtabular}
		\begin{tabular}{lcccc}
			$R_0$ (GeV$^{-1}$) & \multicolumn{4}{c}{Mass (MeV)} \\
			& $M(0^{++})$ & $M(1^{+-})$ & $M(1^{++})$ & $M(2^{++})$ \\
			\hline
			$R_0 = 6$  & 10799, 10764 & 10772, 10818 & 10819 & 10846 \\
			$R_0 = 8$  & 10793, 10758 & 10766, 10812 & 10813 & 10840 \\
			$R_0 = 10$ & 10791, 10756 & 10765, 10810 & 10812 & 10839 \\
			$R_0 = 12$ & 10791, 10756 & 10764, 10810 & 10811 & 10838 \\
		\end{tabular}
		\label{tab:mass_values_bb_sbar_s}
	\end{ruledtabular}
\end{table}

The hidden-bottom tetraquark states can be realized by converting the charm quarks of the hidden-charm tetraquark states into bottom quarks. 
And these hidden-bottom tetraquark states can be classified into three configurations: $ b\bar{b}n\bar{n} $, $ b\bar{b}n\bar{s} $, and $ b\bar{b}s\bar{s} $. 
The possible quantum number assignments for these $S$-wave tetraquark states are the same as for the hidden-charm tetraquark states, i.e., $J^{P(C)}=0^{+(+)}$, $1^{+(\pm)}$, and $2^{+(+)}$.
Similarly, given $R_0 = 10$ GeV$^{-1}$, we calculated the mass spectra of these states and plotted them in Figure \ref{fig:three_images_2}.
In addition, considering the influence of different $R_0$ on the mass spectra of hidden-bottom tetraquark states, we take $R_0 = 6, 8, 10$, and $12$ GeV$^{-1}$ respectively, and the corresponding calculated mass spectra listed in Tables \ref{tab:mass_values_bb_qbar_q}-\ref{tab:mass_values_bb_sbar_s}.
The analytical approach for these states is similar to that used for the hidden-charm tetraquark states. 
Therefore, we will not discuss each state in detail.

In the subsection, we focus on the two exotic hadrons observed experimentally: $ Z_b(10610) $ and $ Z_b(10650) $. 
They are found in the decay channels $\pi^\pm \Upsilon(nS) $ and $\pi^\pm h_b(mP) $, where $ n = 1, 2, 3 $ and $ m = 1, 2 $. 
This suggests that the component configuration of these two exotic hadrons may be a tetraquark state $ b\bar{b}n\bar{n} $.
And their quantum numbers are likely $ J^P = 1^+ $, but there is no information regarding their $ C $-parity. 
In our calculations, we prefer to interpret these two exotic states as the tetraquark states $b\bar{b}d\bar{u}$ or $b\bar{b}u\bar{d}$ with quantum number assignment $I(J^{PC})=1(1^{+-})$, as shown in Figure \ref{fig:three_images_2}(a).
However, the calculated masses of the tetraquark states are smaller than the corresponding experimental values for the two exotic hadrons.
One obvious reason is that we chose a larger parameter $R_0$, see Table \ref{tab:mass_values_bb_qbar_q}.
A deeper reason is that the two quarks inside a hadron form a diquark pair $Qq$, which manifests as an attractive interaction.
Compared with the hidden charm tetraquark system, the parameter $R_0$ corresponding to the hidden bottom tetraquark system should be smaller.

\section{summary \label{S4}}
In this paper, we investigate the hidden heavy flavor tetraquark states $Q\bar{Q}q\bar{q}$ composed of two heavy quarks and two light quarks. A remarkable feature of these tetraquark states is that they conform to a hydrogen-bond-like structure in quantum chromodynamics vertion, which can be treated using the BO approximation. Subsequently, we systematically calculated the mass spectra of $S$-wave tetraquark $Q\bar{Q}q\bar{q}$ system with various quantum numbers ($J^P = 0^{+(+)}$, $1^{+(\pm)}$, and $2^{+(+)}$), and analyzed their decay channels.

Based on the fact that heavy quarks are much heavier than light quarks, we assume that the wave functions of heavy and light quarks in the tetraquark state $Q\bar{Q}q\bar{q}$ can be treated separately. We consider the interaction between light quarks as a perturbation and introduce the BO potential. In specific calculations, the BO potential can be divided into three parts: the QCD Coulomb interaction, the eigenenergy of the light quark systems, and a linear term introduced to account for the color confinement of the tetraquark state. Analogous to the treatment of $H_2$ in QED, we solve the two-body Schrödinger equation to obtain the zeroth-order approximation of the energy occupied by light quarks within the tetraquark state, and thus obtain the eigenvalues of the light quark Hamiltonian. For better comparison, we use X(3872) as input to determine $R_0 = 10 \, \text{GeV}^{-1}$, and study other hidden heavy flavor tetraquark states.

Our calculation results indicate that the exotic hadrons $X(4140)$, $Y(3940)$, $Z_b(10610)$, and $Z_b(10650)$ remain strong candidates for tetraquark states. Besides, we propose that $Z_{cs}(3985)^{-}$, $Z_{cs}(4000)^{+}$, and $Z_{cs}(4123)^{-}$ may be partner states of the tetraquark state $cu\bar{c}\bar{s}/cs\bar{c}\bar{u}$ with quantum number $J^P=1^+$. The $X(3960)$ may be a tetraquark state $cs\bar{c}\bar{s}$ with quantum number $J^{PC}=0^{++}$. In addition, we predict several other possible hidden heavy flavor tetraquark states, and look forward to further experimental and theoretical research. Finally, it is worth emphasizing that although the difference in quark masses may have some impact on the mass spectra of tetraquark states, this hydrogen-bond-like structure in quantum chromodynamics should not be overlooked.

\begin{acknowledgments}
The authors would like to thank Yanmei Xiao and Lei Liu for helpful comments and discussions. This work was supported by the Scientific Research Foundation of Chengdu University of Technology under grant No.10912-KYQD2022-09557.
	
\end{acknowledgments}

\bibliography{boref}

\end{document}